\documentclass{PoS}

\usepackage[T1]{fontenc}
\usepackage{booktabs}
\usepackage[dvips,table]{xcolor} 
\usepackage{xspace}
\usepackage{dcolumn}
\usepackage{caption}
\usepackage
[subrefformat=parens,position=top,skip=-15pt,margin=15pt,justification=justified,singlelinecheck=false]
{subcaption}

\usepackage{amsmath}

\newcommand{\fb}{{\ensuremath\unskip\,\text{fb}}\xspace}
\def\reffi#1{\mbox{Fig.~\ref{#1}}}

\def\refta#1{\mbox{Table~\ref{#1}}}

\def\citere#1{\mbox{Ref.~\cite{#1}}}

\def\be{\begin{equation}}
\def\ee{\end{equation}}

\newcommand{\Pj}{\ensuremath{\text{j}}\xspace}
\newcommand{\Pp}{\ensuremath{\text{p}}\xspace}
\newcommand{\Pe}{\ensuremath{\text{e}}\xspace}

\newcommand{\Pq}{\ensuremath{\text{q}}\xspace}

\newcommand{\Pu}{\ensuremath{\text{u}}\xspace}
\newcommand{\Pd}{\ensuremath{\text{d}}\xspace}

\newcommand{\Pg}{\ensuremath{\text{g}}\xspace}

\newcommand{\PW}{\ensuremath{\text{W}}\xspace}

\newcommand{\TeV}{\ensuremath{\,\text{TeV}}\xspace}

\newcommand{\alphas}{\ensuremath{\alpha_\text{s}}\xspace}

\newcommand{\recola}{{\sc Recola}\xspace}
\newcommand{\mocanlo}{{\sc MoCaNLO}\xspace}
\newcommand{\collier}{{\sc Collier}\xspace}

\newcolumntype{.}{D{.}{.}{-1}}
\newcolumntype{d}[1]{D{.}{.}{#1}}

\colorlet{tableoverheadcolor}{gray!37.5}
\colorlet{tableheadcolor}{gray!25}
\colorlet{tablerowcolor}{gray!12.5}

\newlength{\width}
\newlength{\height}

\def\draftdate{\relax}
\def\mda{\relax}
\def\mua{\relax}
\def\mla{\relax}
\def\draft{
\def\thtystars{******************************}
\def\sixtystars{\thtystars\thtystars}
\typeout{}
\typeout{\sixtystars**}
\typeout{* Draft mode!
         For final version remove \protect\draft\space in source file *}
\typeout{\sixtystars**}
\typeout{}
\def\draftdate{\today}
\def\mua{\marginpar[\boldmath\hfil$\uparrow$]%
                   {\boldmath$\uparrow$\hfil}\color{black}%
                    \typeout{marginpar: $\uparrow$}\ignorespaces}
\def\mda{\color{red}\marginpar[\boldmath\hfil$\downarrow$]%
                   {\boldmath$\downarrow$\hfil}%
                    \typeout{marginpar: $\downarrow$}\ignorespaces}
\def\mla{\marginpar[\boldmath\hfil$\rightarrow$]%
                   {\boldmath$\leftarrow $\hfil}%
                    \typeout{marginpar: $\leftrightarrow$}\ignorespaces}
\def\Mua{\marginpar[\boldmath\hfil$\Uparrow$]%
                   {\boldmath$\Uparrow$\hfil}\color{black}%
                    \typeout{marginpar: $\uparrow$}\ignorespaces}
\def\Mda{\color{red}\marginpar[\boldmath\hfil$\Downarrow$]%
                   {\boldmath$\Downarrow$\hfil}%
                    \typeout{marginpar: $\downarrow$}\ignorespaces}
\def\Mla{\marginpar[\boldmath\hfil\textcolor{red}{$\Rightarrow$}]%
                   {\boldmath\textcolor{red}{$\Leftarrow $}\hfil}%
                    \typeout{marginpar: $\leftrightarrow$}\ignorespaces}
\overfullrule 5pt
\oddsidemargin 15mm
\marginparwidth 29mm
}

\title{Electroweak corrections to vector-boson scattering}

\ShortTitle{Electroweak corrections to vector-boson scattering}

\author{Benedikt Biedermann, \speaker{Ansgar Denner}, Mathieu Pellen\\
        Julius-Maximilians Universit\"at W\"urzburg, W\"urzburg, Germany\\
        E-mail: \email{mathieu.pellen@physik.uni-wuerzburg.de}, \email{ansgar.denner@physik.uni-wuerzburg.de}, \email{mathieu.pellen@physik.uni-wuerzburg.de}}

      \abstract{We report on a recent calculation of the complete NLO
        QCD and electroweak corrections to the process
        $\Pp\Pp\to\mu^+\nu_\mu\Pe^+\nu_{\Pe}\Pj\Pj$, i.e.\ like-sign
        charged vector-boson scattering. The computation is based on
        the complete amplitudes involving two different orders of the
        strong and electroweak coupling constants at tree level and
        three different orders at one-loop level. We find electroweak
        corrections of $-13\%$ for the fiducial cross section that are
        an intrinsic feature of the vector-boson scattering process.
        For differential distributions, the corrections reach up to
        $-40\%$ in the phase-space regions explored. At the NLO level
        a unique separation between vector-boson scattering and
        irreducible background processes is not possible any more at
        the level of Feynman diagrams.}

\FullConference{13th International Symposium on Radiative Corrections (Applications of Quantum Field Theory to
Phenomenology)\\
                 25-29 September, 2017 \\
                 St. Gilgen, Austria}

\begin{document}

\section{Introduction}

The scattering of electroweak (EW) massive vector bosons is a crucial
process to be studied at the Large Hadron Collider (LHC). It allows to
test the Higgs sector of the Standard Model and provides a very
sensitive probe to New Physics. The most promising vector-boson
scattering (VBS) channel is the one where to equally charged W bosons
scatter and subsequently decay into two equally charged leptons and
two neutrinos.
Evidence for VBS in the same-sign WW~channel has been reported based
on run I data \cite{Aad:2014zda,Aaboud:2016ffv,Khachatryan:2014sta},
and the CMS collaboration has observed this process with data from
run~II \cite{CMS:2017adb}.

The complete measurable physical process encompassing same-sign VBS is
given by $\Pp\Pp\to\mu^+\nu_\mu\Pe^+\nu_{\Pe}\Pj\Pj$.  Parts of the
corresponding next-to-leading-order (NLO) corrections have already
been computed in the past, such as the NLO QCD corrections for the
EW-induced process~\cite{Jager:2009xx,Jager:2011ms,Denner:2012dz} and
its QCD-induced irreducible background
\cite{Melia:2010bm,Melia:2011gk,Campanario:2013gea}.

In this proceedings contribution we summarize our calculation
\cite{Biedermann:2017bss} of the complete NLO corrections to the full
process $\Pp\Pp\to\mu^+\nu_\mu\Pe^+\nu_{\Pe}\Pj\Pj$ featuring in
particular surprisingly large NLO EW corrections
\cite{Biedermann:2016yds,Biedermann:2017ozu}.

\section
{Calculating NLO corrections to $\Pp\Pp\to\mu^+\nu_\mu\Pe^+\nu_{\Pe}\Pj\Pj$}

\begin{figure}
\begin{center}
          \includegraphics[width=0.30\linewidth]{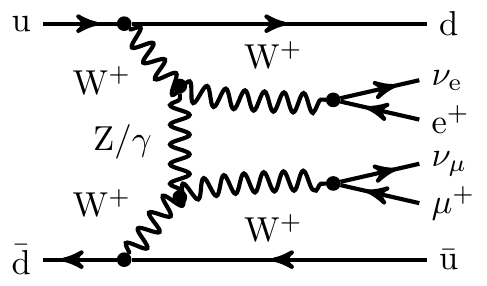}
%          \includegraphics[width=0.30\linewidth]{figures/LO_EW_4}
%          \includegraphics[width=0.30\linewidth]{figures/LO_EW_5}
%          \raisebox{.5ex}{\includegraphics[width=0.35\linewidth]{figures/LO_EW_2}}
          \raisebox{-1.8ex}{\includegraphics[width=0.32\linewidth]{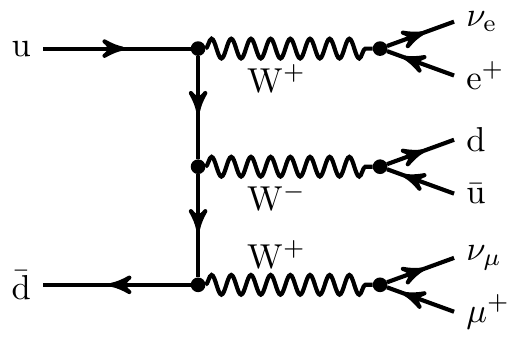}}
          \raisebox{.1ex}{\includegraphics[width=0.30\linewidth]{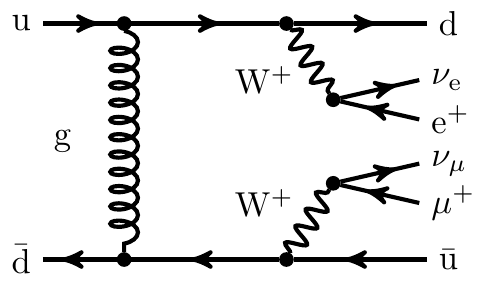}}
\end{center}
        \caption{Sample tree-level diagrams contributing to the
          process
          $\Pp\Pp\to\mu^+\nu_\mu\Pe^+\nu_{\Pe}\Pj\Pj$.}
\label{diag:LO}
\end{figure}

At the amplitude level, the process
$\Pp\Pp\to\mu^+\nu_\mu\Pe^+\nu_{\Pe}\Pj\Pj$ receives two different
types of contributions at leading order (LO), pure EW contributions of order
$\mathcal{O}{\left(g^{6}\right)}$ and QCD-induced contributions of
order $\mathcal{O}{\left(g_{\rm s}^2 g^{4}\right)}$, where $g$ and
$g_{\rm s}$ are the EW and strong coupling constants, respectively.
In \reffi{diag:LO} we show some tree-level diagrams for the partonic
sub-process $\Pu\bar\Pd\to\mu^+\nu_\mu\Pe^+\nu_{\Pe}\bar\Pu\Pd$.  The
$t$-channel diagram on the left illustrates the characteristic VBS
topology of two $\PW$ bosons with space-like momenta that scatter into
two $\PW$~bosons with time-like momenta.  The $s$-channel diagram in
the middle contributes to the irreducible EW background; it actually
features triple gauge-boson production.  Finally, the diagram on
the right-hand side provides an example of a QCD-induced contribution.
Thus, at the level of the cross section three gauge-invariant
contributions exist: the pure EW contribution of order
$\mathcal{O}{\left(\alpha^{6}\right)}$, the QCD-induced contribution
of order $\mathcal{O}{\left(\alpha_{\rm s}^{2}\alpha^{4}\right)}$, and
the interferences of the order
$\mathcal{O}{\left(\alphas\alpha^{5}\right)}$.  Owing to the colour
structure, the interferences occur only in partonic channels that
involve contributions of two different kinematic channels ($s$, $t$,
$u$). In our calculation we include all contributions that belong to
the hadronic process $\Pp\Pp\to\mu^+\nu_\mu\Pe^+\nu_{\Pe}\Pj\Pj$ at
tree and one-loop order.

\begin{figure}
\begin{center}
          \includegraphics[width=0.3\linewidth]{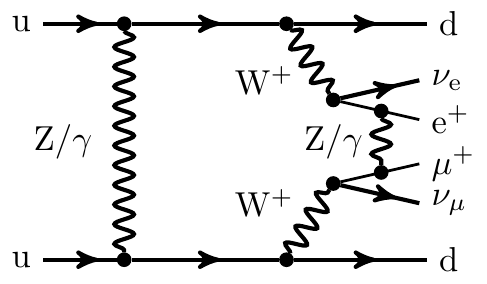}
          \includegraphics[width=0.34\linewidth]{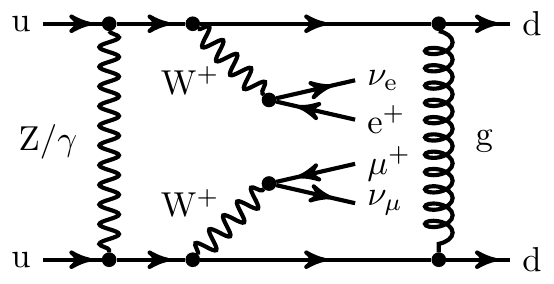}
          \includegraphics[width=0.3\linewidth]{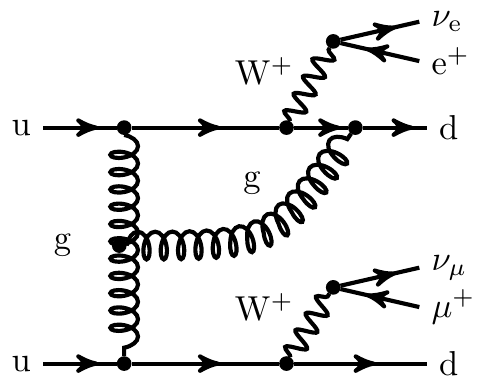}
\end{center}
        \caption{One-loop level diagrams contributing to the process $\Pp\Pp\to\mu^+\nu_\mu\Pe^+\nu_{\Pe}\Pj\Pj$.}
\label{diag:NLO}
\end{figure}

The NLO corrections involve contributions of orders:
$\mathcal{O}{\left(\alpha^{7}\right)}$,
$\mathcal{O}{\left(\alphas\alpha^{6}\right)}$,
$\mathcal{O}{\left(\alphas^{2}\alpha^{5}\right)}$, and
$\mathcal{O}{\left(\alphas^{3}\alpha^{4}\right)}$.  The contributions
of order $\mathcal{O}{\left(\alpha^{7}\right)}$ furnish the NLO EW
corrections to the EW-induced LO process \cite{Biedermann:2016yds}.
The order $\mathcal{O}{\left(\alpha_{\rm s}^{3}\alpha^{4}\right)}$
contributions are the QCD corrections to the QCD-induced process
\cite{Melia:2010bm,Campanario:2013gea}.  For the orders
$\mathcal{O}{\left(\alphas\alpha^{6}\right)}$ and
$\mathcal{O}{\left(\alphas^{2}\alpha^{5}\right)}$, a simple separation
into QCD and EW corrections is not possible any more.  For instance, the order
$\mathcal{O}{\left(\alphas\alpha^{6}\right)}$ contains QCD corrections
to the EW-induced process
\cite{Jager:2009xx,Jager:2011ms,Denner:2012dz,Campanario:2013gea}
as well as EW corrections to the LO interference.  In fact, the
virtual corrections of order $\mathcal{O}{\left(g_{\rm
      s}^{2}g^{6}\right)}$ involve diagrams that can be
interpreted as EW correction to the QCD-induced process or as QCD
correction to the EW-induced process (see diagram in the middle of
\reffi{diag:NLO}). Therefore, both processes cannot be separated any
more once the full NLO corrections are included.

Some contributing one-loop diagrams are shown in \reffi{diag:NLO}.
The diagram on the left-hand side 
of order $\mathcal{O}{\left(g^{8}\right)}$ involves an
8-point function, the one in the middle is  of order
$\mathcal{O}{\left(g_{\rm s}^{2}g^{6}\right)}$ and the one on the
right hand-side of order $\mathcal{O}{\left(g_{\rm s}^{4}g^{4}\right)}$.
All tree-level and one-loop matrix elements have been obtained
with \recola~\cite{Actis:2012qn,Actis:2016mpe} in combination with 
\collier~\cite{Denner:2014gla,Denner:2016kdg}.
Throughout, the complex-mass scheme \cite{Denner:1999gp,Denner:2005fg}
is used.  

To handle the IR divergences (of QCD and QED origin) in the real NLO
corrections the dipole-subtraction method for QCD \cite{Catani:1996vz}
and its extension to QED \cite{Dittmaier:1999mb} have been employed.
The colour-correlated matrix elements needed for the subtraction
procedure are obtained directly from \recola.  The QCD radiation of
type $\Pp\Pp\to\mu^+\nu_\mu\Pe^+\nu_{\Pe}\Pj\Pj\Pj$ includes both
gluon radiation from any coloured particle as well as quark/anti-quark
radiation from contributions with $\Pg\bar{\Pq}$ and $\Pg\Pq$ initial
states.  Similarly, real radiation from photon-induced contributions
of the type $\gamma \Pq / \gamma \bar{\Pq}
\to\mu^+\nu_\mu\Pe^+\nu_{\Pe}\Pj\Pj\Pj$ contributes at the orders
$\mathcal{O}{\left(\alpha^{7}\right)}$,
$\mathcal{O}{\left(\alphas\alpha^{6}\right)}$, and
$\mathcal{O}{\left(\alphas^{2}\alpha^{5}\right)}$.  The photon-induced
corrections are at the level of $2\%$ and originate mainly from the
contributions of order $\mathcal{O}{\left(\alpha^{7}\right)}$.  We
have computed these contributions separately but do not include them
in the results shown below. Results for the  photon-induced
corrections can be found in \citere{Biedermann:2016lvg}.

\section{Numerical results}

The input parameters and the selection cuts for the numerical analysis
can be found in \citere{Biedermann:2017bss} and are inspired by the
experimental searches for the VBS process
\cite{Aad:2014zda,Khachatryan:2014sta,CMS:2017adb}. For the
  centre-of-mass energy we have chosen $13\TeV$.
The renormalisation and factorisation scales are set dynamically to
\begin{equation}
\label{eq:defscale}
 \mu_{\rm ren} = \mu_{\rm fac} = \sqrt{p_{\rm T, j_1}\, p_{\rm T, j_2}},
\end{equation}
where $p_{\rm T, j_i}$, $i=1,2$, are the transverse momenta of the two
jets with largest transverse momenta within the fiducial volume. 
To determine the scale variation,
the central scale \eqref{eq:defscale} has been scaled by factors
$\xi_{\rm fac}$ and $\xi_{\rm ren}$ within the set
\begin{align}
\label{eq:scales}
% (\xi_{\rm fac},\xi_{\rm ren}) \in 
\big\{\left(1/2,1/2\right),\,
 \left(1/2,1\right),\,
 \left(1,1/2\right),\,
 \left(1,1\right),\,
 \left(1,2\right),\,
 \left(2,1\right),\,
 \left(2,2\right)\big\}.
\end{align}
The scale variation is determined by the maximum
and minimum obtained with these seven values, and the central scale
corresponds to $\xi_{\rm fac}=\xi_{\rm ren} = 1$.

The LO fiducial cross section split into the different orders of
coupling constants is presented in \refta{table:LOcrosssection}.
\begin{table}
\begin{center}
\begin{tabular}{|l||c|c|c||c|}
\hline
Order & $\mathcal{O}{\left(\alpha^{6}\right)}$ & $\mathcal{O}{\left(\alphas\alpha^{5}\right)}$ & $\mathcal{O}{\left(\alphas^2\alpha^{4}\right)}$ & Sum \\
\hline
%\hline
${\sigma_{\mathrm{LO}}}$ [fb] 
& $1.4178(2)$
& $0.04815(2)$
& $0.17229(5)$
& $1.6383(2)$ \\
%\hline
%\hline
%$\sigma^{\rm max}_{\mathrm{LO}}$ [fb] 
%& $1.5443(2)$
%& $0.05680(3)$
%& $0.22821(6)$
%& $1.8293(2)$ \\
%\hline
%$\sigma^{\rm min}_{\mathrm{LO}}$ [fb] 
%& $1.3091(2)$
%& $0.04135(2)$
%& $0.13323(3)$
%& $1.4836(2)$ \\
\hline
\end{tabular}
\end{center}
\caption{
Fiducial cross section at LO for the process
$\Pp\Pp\to\mu^+\nu_\mu\Pe^+\nu_{\Pe}\Pj\Pj$ in femtobarn. 
The statistical uncertainty from the Monte Carlo integration on the last digit is given in parenthesis.}
\label{table:LOcrosssection}
\end{table}
For the fiducial volume with VBS cuts the EW-induced process is
clearly dominating over its irreducible background processes amounting
to $87\%$ of the cross section of the full process
$\Pp\Pp\to\mu^+\nu_\mu\Pe^+\nu_{\Pe}\Pj\Pj$. For the scale dependence
of the integrated LO cross section we find
 \begin{equation}
\sigma_{\rm LO} = 1.6383(2)_{-9.44(2)\%}^{+11.66(2)\%}\fb.
\end{equation}

In \refta{table:NLOcrosssection}, all NLO corrections to the
fiducial cross section split into contributions of the 
different orders in the strong and EW coupling are presented. 
\begin{table}
\begin{center}
\begin{tabular}{|l||c|c|c|c||c|}
\hline
Order & $\mathcal{O}{\left(\alpha^{7}\right)}$ & $\mathcal{O}{\left(\alphas\alpha^{6}\right)}$ & $\mathcal{O}{\left(\alphas^{2}\alpha^{5}\right)}$ & $\mathcal{O}{\left(\alphas^{3}\alpha^{4}\right)}$ & Sum \\
\hline
\hline 
${\delta \sigma_{\mathrm{NLO}}}$ [fb] 
& $-0.2169(3)$ 
& $-0.0568(5)$
& $-0.00032(13)$
& $-0.0063(4)$ 
& $-0.2804(7)$ \\
\hline
$\delta \sigma_{\mathrm{NLO}}/\sigma_{\rm LO}$ [\%] & $-13.2$ & $-3.5$ & $0.0$ & $-0.4$ & $-17.1$ \\
\hline
\end{tabular}
\end{center}
\caption{
Absolute and relative NLO corrections for the process
$\Pp\Pp\to\mu^+\nu_\mu\Pe^+\nu_{\Pe}\Pj\Pj$ at the orders  
$\mathcal{O}{\left(\alpha^{7}\right)}$, $\mathcal{O}{\left(\alphas\alpha^{6}\right)}$, $\mathcal{O}{\left(\alphas^{2}\alpha^{5}\right)}$, and $\mathcal{O}{\left(\alphas^{3}\alpha^{4}\right)}$ and for the sum of all NLO corrections.
The statistical uncertainty from the Monte Carlo integration on the last digit is given in parenthesis.}
\label{table:NLOcrosssection}
\end{table}
The total relative NLO corrections, normalised to the sum of all LO
contributions to the full process, amounts to $-17.1\%$.  The bulk of
the correction with $-13.2\%$ originates from the EW corrections of order
$\mathcal{O}{\left(\alpha^{7}\right)}$. The order
$\mathcal{O}{\left(\alphas\alpha^{6}\right)}$ corrections amount to
$-3.5\%$, while the other contributions are below a per cent.  The
scale dependence for the NLO cross section is obtained as
\begin{equation} 
\sigma_{\rm NLO} =  1.3577(7)_{-2.7(1)\%}^{+1.2(1)\%}\fb, 
\end{equation}
corresponding to a reduction by a factor five with respect to the LO.

Nonetheless, the LO and NLO uncertainty intervals do not overlap.  The
reason is that the purely EW contributions dominate the
$\mu^+\nu_\mu\Pe^+\nu_{\Pe}\Pj\Pj$ final state, and the NLO EW
corrections to VBS represent a large fraction of the NLO corrections.
These corrections simply shift the prediction without affecting the
size of the scale variation band.  While missing higher-order QCD
corrections can be estimated via scale variations this is not the case
for higher-order EW corrections in the on-shell scheme. A conservative
estimate for the higher-order EW corrections is provided by the square
of the relative EW NLO correction.

The origin of the large EW corrections to the fiducial cross section
has been elucidated in \citere{Biedermann:2016yds}. Using the
double-pole approximation and the effective vector-boson approximation
for the matrix elements and in addition the high-energy logarithmic
approximation for the EW corrections, the large EW corrections can be
reproduced. This analysis showed that the EW corrections are large in
comparison to those of other LHC processes owing to the relatively
high intrinsic energy scale of VBS, the large EW charges of the
W~bosons, and the relatively weak cancellation between leading and
subleading EW logarithms.

\begin{figure}
        \setlength{\parskip}{-10pt}
        \begin{subfigure}{0.49\textwidth}
                \subcaption{}
                \includegraphics[width=\textwidth]{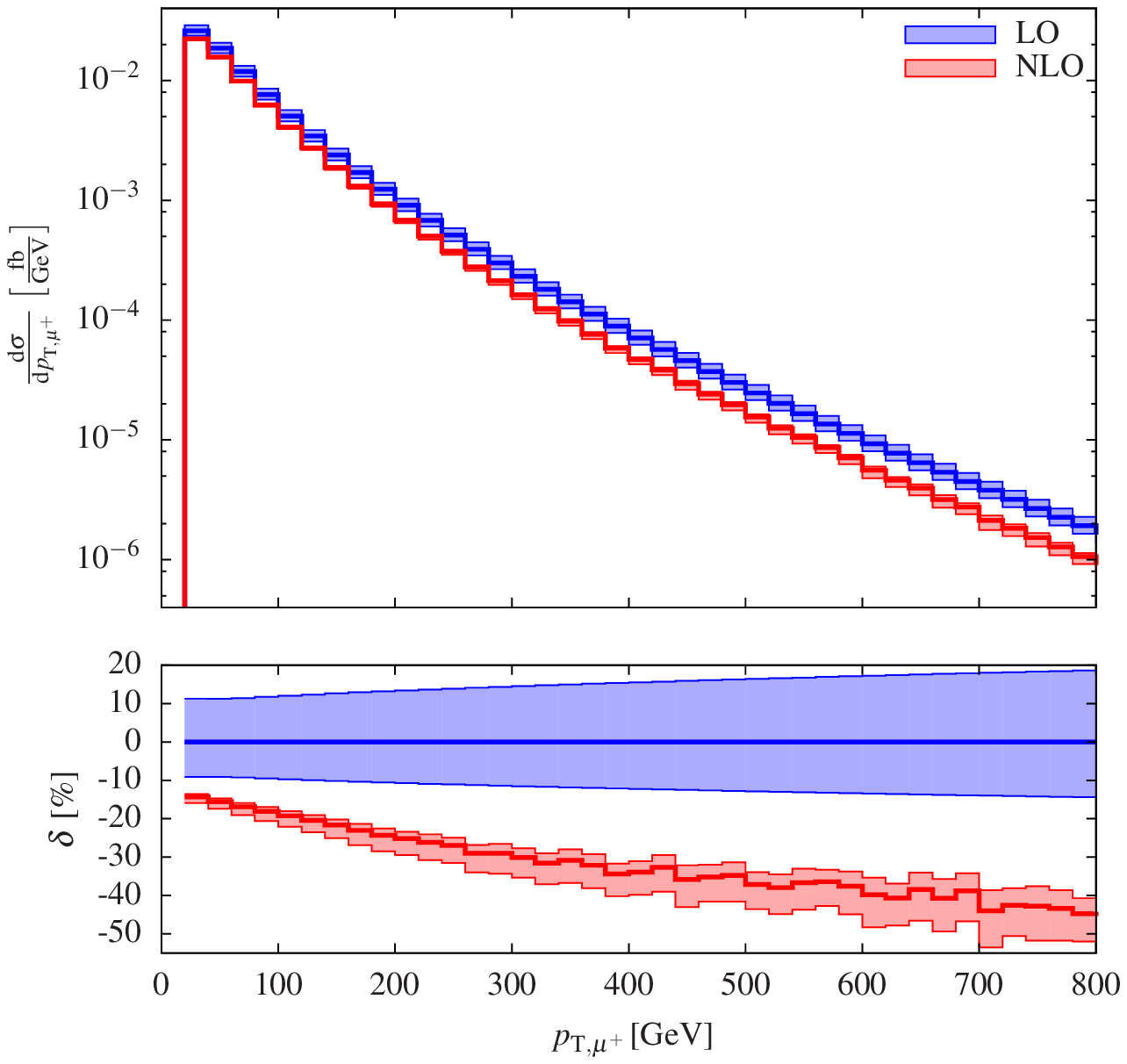}
                \label{plot:transverse_momentum_antimuon_scale}
        \end{subfigure}
        \hfill
        \begin{subfigure}{0.49\textwidth}
                \subcaption{}
                \includegraphics[width=\textwidth]{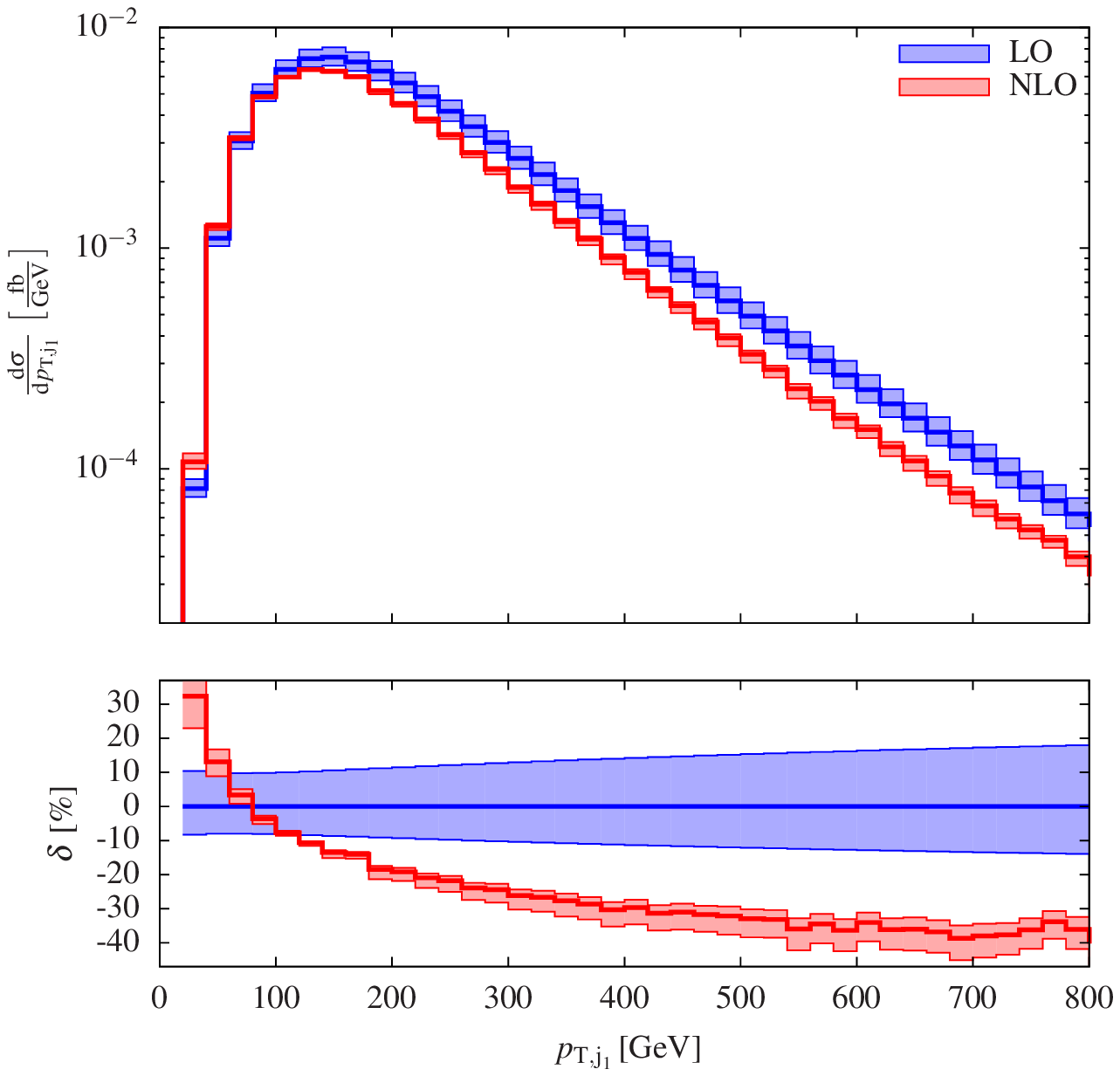}
                \label{plot:histogram_transverse_momentum_j1_scale}
        \end{subfigure}
        
        \vspace*{-3ex}
        \caption{\label{fig:distributions_scale}%
                Differential distributions at a centre-of-mass energy $\sqrt{s}=13\TeV$ at the LHC for $\Pp\Pp\to\mu^+\nu_\mu\Pe^+\nu_{\Pe}\Pj\Pj$: 
                \subref{plot:transverse_momentum_antimuon_scale}~transverse momentum of the anti-muon~(left), 
                \subref{plot:histogram_transverse_momentum_j1_scale}~transverse momentum of the hardest jet~(right).
                The upper panels show the sum of all LO and NLO contributions with scale variation.
                The lower panels show the relative corrections
                in per cent.  }
\end{figure}

The contributions to different coupling orders in the NLO corrections
to $\Pp\Pp\to\mu^+\nu_\mu\Pe^+\nu_{\Pe}\Pj\Pj$ as well as the
photon-induced contributions to various distributions have been
discussed in \citere{Biedermann:2017bss}. Here, we merely show some
additional results for the dependence of the complete NLO corrections on
the factorisation and renormalisation scales in
\reffi{fig:distributions_scale}. Corresponding results for other
distributions can be found in  \citere{Biedermann:2017bss}.
In the upper panels, the complete LO prediction as well as the
complete NLO prediction are shown.  The band is obtained by varying
the factorisation and renormalisation scales independently by the
factors $\xi_{\rm fac}$ and $\xi_{\rm ren}$ with the combinations of
\eqref{eq:scales}.  The complete relative NLO corrections shown in the
lower panel are normalised to the LO prediction for the central scale.
In \reffi{plot:transverse_momentum_antimuon_scale}, the distribution
in the transverse momentum of the anti-muon is displayed, and the
distribution in the transverse momentum of the hardest jet is shown in
\reffi{plot:histogram_transverse_momentum_j1_scale}.  The
scale-uncertainty band decreases significantly by going from LO to
NLO for all distributions.
The NLO corrections are larger for 
distributions in leptonic transverse momenta or invariant masses 
than for those in jet transverse momenta or invariant masses.  In
general, the scale dependence is larger where the cross section is
smaller and the NLO corrections are larger.

\section{Conclusions}
\label{sec:Conclusions}

In this proceedings contribution we have reported on the first
calculation of the complete NLO EW and QCD corrections to the process
$\Pp\Pp\to\mu^+\nu_\mu\Pe^+\nu_{\Pe}\Pj\Pj$ including like-sign
charged vector-boson scattering (VBS) and its EW- and QCD-induced
irreducible background. Using \recola, the full LO and NLO matrix
elements are used throughout the calculation, including possible
off-shell, non-resonant, and interference effects.  While at LO the
EW-induced and the QCD-induced contributions can be unambiguously
separated based on the Feynman diagrams and coupling constants, this
is not possible anymore at NLO.  Hence, for the full process at NLO
the VBS process cannot strictly be distinguished from its irreducible
background.

The NLO corrections are dominated by large negative EW corrections.
For the fiducial cross section, they reach $-13\%$ of the complete LO
contributions and are even significantly more enhanced at the level of
differential distributions with up to (minus) $40\%$ corrections in
the kinematical regions explored.  These corrections display the
typical behaviour of Sudakov logarithms that grow large in the
high-energy regime.  The dependence on the factorisation and
renormalisation scale is significantly reduced upon including NLO
corrections.  However, this does not provide an estimate of the
theoretical uncertainty from missing higher-order EW corrections.

\acknowledgments We thank J.-N.~Lang and S.~Uccirati for
supporting the computer program \recola\ and R.~Feger for
assistance with the Monte Carlo program \mocanlo.  We acknowledge
financial support by the German Federal Ministry for Education and
Research (BMBF) under contract no.~05H15WWCA1 and the German Science
Foundation (DFG) under reference DE 623/6-1.

\bibliographystyle{JHEPmod}
\bibliography{vbs_nlo} 

%\begin{thebibliography}{99}
% \bibitem{...} ....
%\end{thebibliography}

\end{document}